\documentclass[fleqn,usenatbib]{mnras}

\usepackage[T1]{fontenc}

\DeclareRobustCommand{\VAN}[3]{#2}
\let\VANthebibliography\thebibliography
\def\thebibliography{\DeclareRobustCommand{\VAN}[3]{##3}\VANthebibliography}

\usepackage{graphicx}	
\usepackage{amsmath}	
\usepackage{tabularx}
\usepackage{mathtools}
\usepackage{graphicx}
\usepackage{amssymb}
\usepackage{latexsym}
\usepackage{amsmath}
\usepackage{newtxtext,newtxmath}

\DeclarePairedDelimiter{\abs}{\lvert}{\rvert}

\title[Dynamical history of the Galilean satellites for a fast migration of Callisto]{Dynamical history of the Galilean satellites for a fast migration of Callisto}

\author[Giacomo Lari, Melaine Saillenfest, and Clara Grassi]{
Giacomo Lari,$^{1}$\thanks{E-mail: giacomo.lari@unipi.it}
Melaine Saillenfest,$^{2}$
Clara Grassi$^{1}$
\\
$^{1}$Dipartimento di Matematica, Universit\`a di Pisa, Largo Bruno Pontecorvo 5, 56127 Pisa, Italy\\
$^{2}$IMCCE, Observatoire de Paris, PSL Research University, CNRS, Sorbonne Universit\'e, Universit\'e de Lille, 75014 Paris, France
}

\date{Accepted XXX. Received YYY; in original form ZZZ}

\pubyear{2022}

\begin{document}
\label{firstpage}
\pagerange{\pageref{firstpage}--\pageref{lastpage}}
\maketitle

\begin{abstract}
The dynamics of the innermost Galilean satellites (Io, Europa and Ganymede) is characterised by a chain of mean motion resonances, called Laplace resonance, and by a strong tidal dissipation that causes wide variations of their semi-major axes over large timescales. The precise history of energy dissipation in the Jovian system is not known, but several theories have been proposed. Tidal resonance locking states that big outer moons can also migrate fast. If this is the case for Callisto, then it should have crossed the 2:1 mean motion resonance with Ganymede in the past, affecting the motion of all four Galilean satellites. Therefore, we aim to determine whether a fast migration for Callisto is compatible with the current orbital configuration of the system. Due to the chaotic nature of the resonant crossing, different outcomes are possible. A small portion of our simulations shows that Callisto can cross the 2:1 resonance with Ganymede without being captured and preserving the Laplace resonance. However, in most cases, we found that Callisto is captured into resonance, despite its divergent migration. As Callisto continues to migrate fast outwards, the moons depart substantially from the exact 8:4:2:1 commensurability, while still maintaining the resonant chain. Callisto can eventually escape it by crossing a high-order mean motion resonance with Ganymede. Afterwards, the moons' system is able to relax to its current configuration for suitable dissipation parameters of the satellites. Therefore it is possible, although challenging, to build a self-consistent picture of the past history of the Galilean satellites for a fast migration of Callisto.
\end{abstract}




\section{Introduction}

The Galilean satellites are by far the largest moons of Jupiter: in order from the planet, they are Io (1), Europa (2), Ganymede (3) and Callisto (4). While Io is characterised by strong volcanism all over its surface and does not show evidence of water, the other three moons are covered by ice and probably conceal vast oceans of liquid water under their surfaces (e.g. \citealp{SCHUBERT-etal_2004}). Because of their peculiar characteristics, the Galilean satellites have been extensively studied in the last decades and two large missions, Europa Clipper by NASA and JUICE by ESA, are going to visit them and collect a great variety of observations. Moreover, apart from their interior and surface features, the Galilean satellites are noteworthy also for their orbital configuration. Indeed, the three inner moons are locked in a chain of mean motion resonances (MMRs) composed by a 2:1 resonance between Io and Europa, and a 2:1 resonance between Europa and Ganymede. More precisely, writing $\lambda_i$ the mean longitude of the $i$th satellite and $\varpi_i$ its longitude of pericentre, we currently have
\begin{equation}\label{lap2bod}
   \begin{aligned}
      \lambda_1-2\lambda_2+\varpi_1&\sim 0 \,,\\
      \lambda_1-2\lambda_2+\varpi_2&\sim \pi \,,\\
      \lambda_2-2\lambda_3+\varpi_2&\sim 0 \,,
   \end{aligned}
\end{equation}
where symbol $\sim$ stands for ``closely oscillates around''. From the last two relations, we obtain:
\begin{equation}\label{lapang}
   \lambda_1-3\lambda_2+2\lambda_3\sim \pi \,,
\end{equation}
which involves the mean longitudes of all three satellites. This relation is commonly known as ``Laplace resonance''.

The origin of this configuration has been investigated since the second half of the twentieth century. There are two main theories that try to explain it. The first one proposes that the resonance was formed through successive resonant captures driven by the tidal dissipation within Jupiter \citep{YODER_1979,YODER-PEALE_1981}: as Io is closer to the planet, it migrated faster than the other satellites and reached the 2:1 resonance with Europa; then the two inner moons moved outwards and captured also Ganymede. The second theory, instead, proposes that the resonance was triggered at the time of the formation of the satellites \citep{GREENBERG_1982,PEALE-LEE_2002}: the moons underwent a Type-I migration inside the circumplanetary disk of Jupiter, and as Ganymede is more massive, it moved inwards faster than the other moons and captured rapidly first Europa and then Io. A primordial origin of the Laplace resonance is also supported by recent formation models of the Galilean satellites \citep{SHIBAIKE-etal_2019,BATYGIN-MORBIDELLI_2020}.

\begin{table}
\caption{Mean orbital elements of the Galilean satellites at J2000 epoch obtained from moons' ephemerides Jup310 released by JPL. Semi-major axes are given in Jupiter's radii, where we set $R_\text{J}=71398$ km.}
{\small
\begin{center}
\noindent\begin{tabular}{r|rrrr}
\hline
element & Io & Europa & Ganymede & Callisto \\
\hline
$a$  & $5.9191$ & $9.4147$ & $15.0157$ & $26.4117$ \\
$e$  & $0.0041$ & $0.0095$ & $0.0015$  & $0.0074$ \\
\hline
  \end{tabular}
\end{center}
}
\label{tab:j2000}
\end{table}

Other works have also investigated the role of past MMRs between the satellites in sculpting the system \citep{TITTEMORE_1990,MALHOTRA_1991}. In general, exploring the past evolution of the Galilean satellites is quite a difficult task, because their current configuration imposes tight constraints (see Eq.~\ref{lap2bod} and Table~\ref{tab:j2000}). We also have an estimate of their migration rate: from astrometric observations, \citet{LAINEY-etal_2009} found that Io is currently moving inwards at about $0.4$ cm/yr, while Europa and Ganymede move outwards at $2$ cm/yr and $11$ cm/yr, respectively. The measured inward migration of Io, that seems to suggest a future breaking of the Laplace resonance, is actually temporary and in the long run all three satellites are expected to move outwards, preserving the Laplace resonance \citep{LARI-etal_2020,CELLETTI-etal_2022}.

\begin{figure}
   \centering
   \includegraphics[scale=0.65]{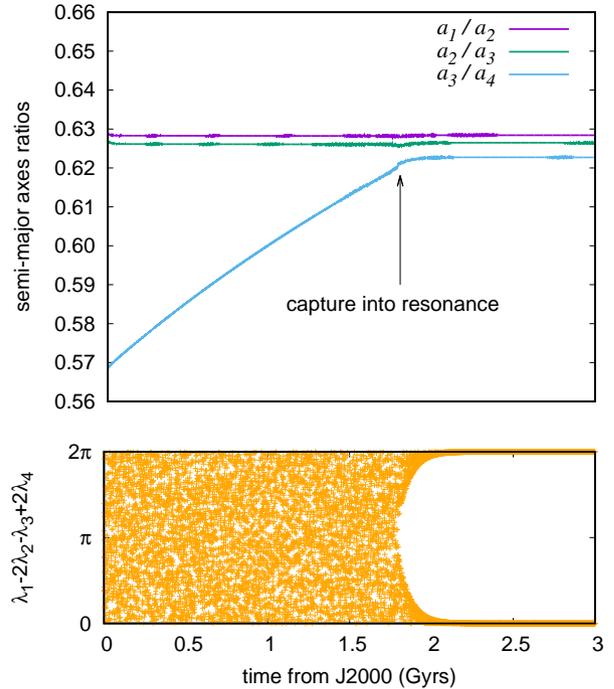}
   \caption{Future evolution of the Galilean satellites as described by \citet{LARI-etal_2020}, assuming that tidal effects at the distance of Callisto are negligible. After about 1.5 Gyr from J2000 epoch, Callisto is captured into a four-body resonant chain with Io, Europa and Ganymede.}
   \label{fig:evolconv}
\end{figure}

\citet{LARI-etal_2020} have investigated the future evolution of the Galilean satellites considering the outward migration of the three inner moons due to the strong tidal dissipation between Jupiter and Io. In their scenario, Callisto's semi-major axis was assumed steady, since, in the classic theory, tidal effects of Jupiter are extremely small at the distance of Callisto. They found that, after about 1.5 billions of years from now, Ganymede will reach the 2:1 resonance with Callisto. As in this case $a_3/a_4$ increases, the encounter is convergent and the two satellites can be captured into a 2:1 MMR (e.g. \citealp{MURRAY-DERMOTT_2000}). Indeed, starting from the current configuration of the system and performing hundreds of simulations, \citet{LARI-etal_2020} found a $100\%$ probability of capture for Callisto, forming a 8:4:2:1 resonant chain with the other three moons (see Fig.~\ref{fig:evolconv}). This chain can be composed of successive 2:1 MMRs between adjacent satellites, or it can involve a pure three-body MMR, which can cause a large increase of the eccentricities of the outer satellites (see also \citealt{MALHOTRA_1991,SHOWMAN-MALHOTRA_1997}).

Recently, \citet{FULLER-etal_2016} presented a new tidal theory called resonance locking, which greatly changes the classic paradigm of tidal effects. According to their theory, dissipation within the planet can increase by several orders of magnitude because of a resonant coupling between a wave in the planet's interior and a satellite moving at the same frequency. In this context, the tidal quality factor $Q$ of gas giants can be far smaller than previously thought \citep{GOLDREICH-SOTER_1966}. Large dissipation in the planet restricted at certain frequencies opens the possibility for a fast tidal migration even for distant satellites. Indeed, this theory provides a natural explanation for the unexpected rate at which Titan is moving away from Saturn, which has been recently measured through the analysis of Cassini radio-science data and astrometric observations \citep{LAINEY-etal_2020}. In light of the resonance locking theory and its successful application to Titan and other satellites of Saturn, Callisto could migrate faster than the other Galilean moons \citep{FULLER-etal_2016}. In this case, the encounter of the 2:1 MMR between Ganymede and Callisto studied by \citet{LARI-etal_2020} may actually have already happened in the past. This scenario matches the formation model presented by \citet{SHIBAIKE-etal_2019}, who find that either all four Galilean satellites formed in a 2:1 MMRs chain, or Callisto formed inside the 2:1 MMR with Ganymede. However, no complete dynamical investigation has been carried out yet, so it is not clear whether a fast migration of Callisto is compatible or not with the current orbital configuration of the system.

Past crossings of MMRs between the moons are possible if their tidal migration timescales $t_{i}^{\text{tide}}=a_i/\dot a_i$ have different values, so that the ratios of their semi-major axes change with time. However, tidal resonance locking can force different moons to migrate at a similar timescale, fixed by the evolution of the interior of the hosting planet; this would prevent the crossing of resonances. Indeed, this is what has been observed for many moons of Saturn, including Titan \citep{LAINEY-etal_2020,CRIDA_2020}. For Jupiter's system, we have $t_{i}^{\text{tide}}\approx 20$ Gyrs for $i=1,2,3$ \citep{LAINEY-etal_2009,FULLER-etal_2016}, while $t_{4}^{\text{tide}}$ is unknown. \citet{FULLER-etal_2016} proposed a current value of $t_{4}^{\text{tide}}$ close to $2$ Gyrs, which would imply that Callisto migrates much faster than the three inner Galilean satellites.

Some recent studies have explored this new dynamical scenario. In particular, \citet{DOWNEY-etal_2020} assumed a very large dissipation within Jupiter at the frequency of Callisto, as proposed by \citet{FULLER-etal_2016}, and considered the orbital variations of Callisto and Ganymede due to the crossing of past MMRs between the two satellites. They computed the optimal values of the moons' dissipative parameters in order to retrieve the current values of their eccentricities and inclinations. However, in their study, \citet{DOWNEY-etal_2020} did not propagate the $N$-body dynamics of the Galilean satellites. Instead, they used simplified analytical expressions for computing the jumps of the orbital elements at each resonance crossing and their damping due to the tidal friction within the moons. By doing so, the authors assumed that Callisto has never been captured into resonance with Ganymede and that the Laplace resonance has never been affected by these resonant encounters. This assumption comes from the fact that two isolated satellites cannot be captured into a two-body MMR if the resonance encounter is divergent (e.g. \citealp{MURRAY-DERMOTT_2000}). However, Ganymede and Callisto are not isolated satellites: Ganymede is already trapped in the Laplace resonance with Io and Europa, and a jungle of three-body resonances surrounds the 2:1 resonance between Ganymede and Callisto, influencing greatly their dynamics (see \citealp{LARI-etal_2020}). Therefore, the outcome of the resonant encounter in this case is much more complicated than what one can predict from isolated two-body resonances. Because of the chaotic nature of the dynamics, the exact outcome of the encounter must be investigated numerically.

Similarly to \citet{SHIBAIKE-etal_2019}, \citet{MADEIRA-etal_2021} have recently proposed that the four Galilean satellites could have triggered a four-body resonance chain just after their formation through their inward migration within Jupiter's circumplanetary disk; then, after a few millions of years, Callisto would have escaped it. In their scenario, Callisto is supposed to be in resonance locking, so that its fast migration allows it to move away from the 2:1 resonance with Ganymede. However, in order to retrieve the current values of the moons' semi-major axes at the end of their simulations, \citet{MADEIRA-etal_2021} had to assume a migration rate for Callisto much smaller than the one predicted by \citet{FULLER-etal_2016} and used by \citet{DOWNEY-etal_2020}.

Starting from the dynamical models described by \citet{LARI_2018} and \citet{LARI-etal_2020}, we aim to explore the dynamical evolution of the Galilean satellites in the case Callisto is in resonance locking and migrating quickly outwards. In such a scenario, Callisto must have crossed the 2:1 MMR with Ganymede in the past. As this crossing must have happened quite recently (less than $1$ Gyr, see \citealp{DOWNEY-etal_2020}), we assume that the Laplace resonance between the three inner satellites was already established at the time, which is the case if it has a primordial origin. 

The paper is structured as follows: in Sect.~\ref{sec:dyn}, we present the dynamical models and the setup we use for investigating the past evolution of the moons. In Sect.~\ref{sec:res}, we show the possible outcomes of the resonant encounter and, in Sect.~\ref{sec:pat}, we study the particular dynamical pathways that involve the formation of a four-body resonant chain and eventually lead to the current configuration of the moons. Finally, in Sect.~\ref{sec:con}, we summarise our results and we comment on possible future advances in our knowledge of tidal dissipation in the Jovian system.

\section{Dynamical models}
\label{sec:dyn}

The crossing of the 2:1 resonance between Callisto and Ganymede is chaotic and it can lead to many different outcomes (see e.g. \citealp{LARI-etal_2020} for examples with convergent migration). Hence, it is not enough to predict the effect of resonances using classic analytical formulas; instead, we must draw a statistical picture of the possible evolution pathways of the moons by running hundreds of simulations over a billion years timespan. For this exploration to be computationally feasible, we need an efficient averaged model that accurately captures the essence of the dynamics of all four moons in the vicinity of the 2:1 resonance between Callisto and Ganymede. Such a model has been introduced by \citet{LARI-etal_2020}. In addition to be light from a computational point of view, this model highlights the limited number of terms that matter most in the dynamics, which allows for an immediate interpretation of the moons' motion.

However, the migration of Callisto predicted by \citet{FULLER-etal_2016} is so fast that the moons can cross many other resonances after having encountered the main 2:1 period ratio. For this reason, this first statistical study must be completed by a limited number of integrations using an un-averaged $N$-body model (that includes all resonances). This way, we can determine which other resonances may also have played a role in the subsequent evolution of the moons, and how they could possibly have led them to their current orbital configuration.

More precisely, we consider that an evolution matches qualitatively the current orbital configuration of the Galilean satellites if at a certain time: i) it reaches the same orbital proportions $a_i/a_{i+1}\ (i=1,2,3)$ as in Table~\ref{tab:j2000}; ii) the only active resonances are the ones described in Eq.~\eqref{lap2bod}; iii) the eccentricity of the inner moons are equal to the values forced by the resonances; iv) Callisto has a moderate free eccentricity ($\lesssim 0.01$).

\subsection{Averaged model}
\label{subsec:avg}

Our averaged model is based on an expansion of the Hamiltonian function that describes the motion of the Galilean moons into series of eccentricities (see \citealt{YODER-PEALE_1981,MALHOTRA_1991,LARI_2018,PAITA-etal_2018}). The details for the construction of the model have been discussed in depth by \citet{LARI_2018} and here we just present the final version of the Hamiltonian. As inclinations and nodes have been proved to play a marginal role in the current Laplace resonance dynamics and in the 2:1 resonant encounter between Ganymede and Callisto \citep{LARI-etal_2020}, we limit our study in Jupiter's equatorial plane. Therefore, minor effects taken into account in \citet{LARI-etal_2020}, such as the Sun's perturbation and inertial forces related to the motion of the equatorial plane of Jupiter, are neglected in this study. This is a reasonable approximation, as the obliquity of Jupiter is small, so the equilibrium plane of the satellites (Laplace plane) is very close to the equator of Jupiter.

The choice of considering a coplanar motion for the satellites is also functional to have a model as simple as possible \citep{YODER_1979,HENRARD_1983,TITTEMORE_1990,MALHOTRA_1991}. However, it is important to note that we are neglecting the constraints to the evolution given by the small (but not zero) current inclinations of the moons.

Following the notation of \citet{LARI-etal_2020}, we denote with $\mathcal{G}$, $m_0$, $R_\text{J}$, and $J_2$ the gravitational constant, the mass, equatorial radius, and quadrupole moment of Jupiter, respectively; while $m_i$, $R_i$ and $\beta_i=m_0m_i/(m_0+m_i)$ $(i=1,2,3,4)$ are the Galilean satellites' masses, radii and reduced masses. Moreover, we use the Keplerian elements of the moons $(a_i,e_i,\varpi_i,\lambda_i)$, which are the semi-major axis of moon $i$, its eccentricity, its longitude of the pericentre, and its mean longitude, respectively. We introduce also the mean motion $n_i=\sqrt{\mathcal{G}(m_0+m_i)/a_i^3}$. We note that in order to keep the Hamiltonian formalism, the actual variables we use for the orbit propagation are not the Keplerian elements, but some combinations of them which define canonical coordinates (for details, see \citealp{LARI-etal_2020}).

The Hamiltonian can be expressed as
\begin{equation}\label{eq:H}
   \mathcal{H} = \mathcal{H}_0 + \varepsilon\mathcal{H}_1 \,,
\end{equation}
where the unperturbed part is a sum of two-body Hamiltonian functions:
\begin{equation}
   \mathcal{H}_0 = -\sum_{i=1}^4 \frac{\mathcal{G}m_0m_i}{2a_i} \,,
\end{equation}
and the perturbation can be decomposed into
\begin{equation}\label{eq:eH1}
   \varepsilon\mathcal{H}_1 = \mathcal{H}_\text{J} + \mathcal{H}_\text{M} \,.
\end{equation}

In this expression, $\mathcal{H}_\text{J}$ is the Hamiltonian function associated to the oblateness of Jupiter. We truncate its expansion to the second order in eccentricities and to the second order in $R_\text{J}/a_i$, and we remove the short period terms, obtaining
\begin{equation}
      \mathcal{H}_\text{J} = \sum_{i=1}^4\frac{\mathcal{G}m_0m_i}{a_i}\Bigg[J_2\left(\frac{R_\text{J}}{a_i}\right)^2\left(-\frac{1}{2} - \frac{3}{4}e_i^2\right)\Bigg] \,.
\end{equation}
The main effect of Jupiter's $J_2$ on the orbits of the satellites is to induce a precession of their longitudes of the pericentre. As the rate of the precession depends on the distance of the moons from the planet, $J_2$ separates the MMRs into multiplets of resonances with the same combination of mean longitudes.

The second term in Eq.~\eqref{eq:eH1} is the Hamiltonian function associated to the mutual gravitational perturbation between the satellites. It can be divided into secular and resonant parts:
\begin{equation}
   \mathcal{H}_\text{M} = \mathcal{H}_\text{M}^{(\text{sec})} + \mathcal{H}_\text{M}^{(\text{res})} \,.
\end{equation}

At second order in eccentricities, the secular part is
\begin{equation}
   \begin{aligned}
      \mathcal{H}_\text{M}^{(\text{sec})} = -\sum_{1\leqslant i<j\leqslant 4} \frac{\mathcal{G}m_im_j}{a_j}\Bigg(&
      f_1 + f_2(e_i^2+e_j^2) \\
      &+ f_{10}\ \ e_ie_j\ \ \cos(\varpi_j-\varpi_i)\Bigg)\,,
   \end{aligned}
\end{equation}
and the resonant part is
\begin{equation}\label{eq:HMres}
   \begin{aligned}
      \mathcal{H}_\text{M}^{(\text{res})} = \!\!\!\!\!\!\!\sum_{ij=(12,23,34)}\!\!\bigg[\frac{\beta_in_ia_i\,\beta_jn_ja_j}{m_0} && e_j & \cos(2\lambda_j -  \lambda_i -  \varpi_j)\\
      -\frac{\mathcal{G}m_im_j}{a_j}\Bigg(
        f_{27}            && e_i    & \cos(2\lambda_j -  \lambda_i -  \varpi_i) \\
      + f_{31}            && e_j    & \cos(2\lambda_j -  \lambda_i -  \varpi_j) \\
      + f_{45}            && e_i^2  & \cos(4\lambda_j - 2\lambda_i - 2\varpi_i) \\
      + f_{53}            && e_j^2  & \cos(4\lambda_j - 2\lambda_i - 2\varpi_j) \\
      + f_{49}            && e_ie_j & \cos(4\lambda_j - 2\lambda_i - \varpi_i - \varpi_j)
      \Bigg)\Bigg]\,,
   \end{aligned}
\end{equation}
where the coefficients $f_k$ are combinations of Laplace coefficients and depend on the ratio of the semi-major axes $a_i/a_j$. Their expression is given, for example, in \citet{MURRAY-DERMOTT_2000}. The resonant part in Eq.~\eqref{eq:HMres} includes the 2:1 resonant terms for all three pairs of adjacent moons. All these terms must be kept in order to study the 2:1 resonance crossing of Ganymede and Callisto while the three inner moons are locked in the Laplace resonance.

From the Hamiltonian in Eq.~\eqref{eq:H}, we can then compute the differential equations that approximate the long-term planar motion of the satellites.

\subsection{Tidal effects}
\label{subsec:tid}

Together with the conservative terms of the dynamics presented in Sect.~\ref{subsec:avg}, we need to include dissipative effects coming from the tidal interaction between Jupiter and the moons. Since these effects cannot be obtained from a Hamiltonian function, we must add them directly to the differential equations. This method is valid as long as the timescale of the dissipative effects is many orders of magnitude greater than the one of the conservative system. Here, indeed, the resonant and secular dynamics of the Galilean satellites have characteristic times from a few years to thousands of years (see e.g. \citealp{LAINEY-etal_2006}), while dissipation acts over millions of years.

The force applied on a given body due to the tidal dissipation is (\citealt{MIGNARD_1979}, see also \citealt{EFROIMSKY-LAINEY_2007,LARI_2018}):
\begin{equation}
   \label{eq:tidemignard}
   \mathbf F=-3\frac{k_2Gm^2R^5}{r^7}\Delta t\left(2\frac{\mathbf{r}}{r}\frac{\mathbf{r}\cdot\mathbf{v}}{r^2}+\frac{\mathbf r \times \mathbf{w} + \mathbf v}{r}\right),
\end{equation}
where $m$, $\mathbf r$ and $\mathbf v$ are the mass, position and velocity of the body that raises the tides, while $k_2$, $R$ and $\mathbf{w}$ are the Love number, radius and spin vector of the deformed body. The parameter $\Delta t$ is called tidal time lag: for tidal dissipation within the planet, $\Delta t=(2(\abs{\mathbf{w}}-n)Q)^{-1}$; instead, for tidal dissipation within the (synchronous) satellite, $\Delta t=(nQ)^{-1}$. The final tidal force applied to the satellite is the sum of both contributions.

In order to implement this force in our averaged model, we need to compute the mean effect of this force on the moons' orbital elements. At lowest order in eccentricities, the dissipative effects on the moons' orbital elements are given by (see \citealt{KAULA_1964,YODER-PEALE_1981,MALHOTRA_1991})
\begin{align}
   \label{eq:dissa}
   \dot{a}_i &= \frac{2}{3}c_i\left(1-\left(7D_i-\frac{51}{4}\right)e_i^2\right)a_i\,, \\
   \label{eq:disse}
   \dot{e}_i & =-\frac{1}{3}c_i\left(7D_i-\frac{19}{4}\right)e_i\,;
\end{align}
where
\begin{align}
\label{eq:cmalho}
      c_i&=\frac{9}{2}\left(\frac{k_2}{Q}\right)_{0,i}\frac{m_i}{m_0}\left(\frac{R_\mathrm{J}}{a_i}\right)^5n_i\,,\\
\label{eq:Dmalho}
      D_i&=\left(\frac{k_2}{Q}\right)_i\left(\frac{Q}{k_2}\right)_{0,i}\left(\frac{R_i}{R_\mathrm{J}}\right)^5\left(\frac{m_0}{m_i}\right)^2\,.
\end{align}

In Eq.~\eqref{eq:cmalho} and~\eqref{eq:Dmalho}, $(k_2/Q)_{0,i}$ is the dissipative parameter of Jupiter at the orbital frequency of the $i$th satellite, while $(k_2/Q)_i$ is the dissipative parameter of the $i$th satellite. These parameters are defined as the ratio between the tidal Love number $k_2$ and the quality factor $Q$. Therefore, in Eqs.~\eqref{eq:dissa} and~\eqref{eq:disse}, the terms proportional to $c_i$ are due to tidal dissipation within Jupiter, while the ones proportional to $c_iD_i$ are due to the tidal dissipation within the $i$th moon. The former pushes the satellite outwards and increases its eccentricity, while the latter pushes the satellite inwards and tends to circularise its orbit. In the case of the Galilean moons, \citet{LAINEY-etal_2009} measured the current dissipative parameters related to the couple Io-Jupiter: $(k_2/Q)_{(0,1)}=1.1\times 10^{-5}$ and $(k_2/Q)_1=0.015$. As we consider a planar motion and inclinations are expected to remain small during the actual evolution, we do not include obliquity tides.

\begin{figure}
   \centering
   \includegraphics[scale=0.65]{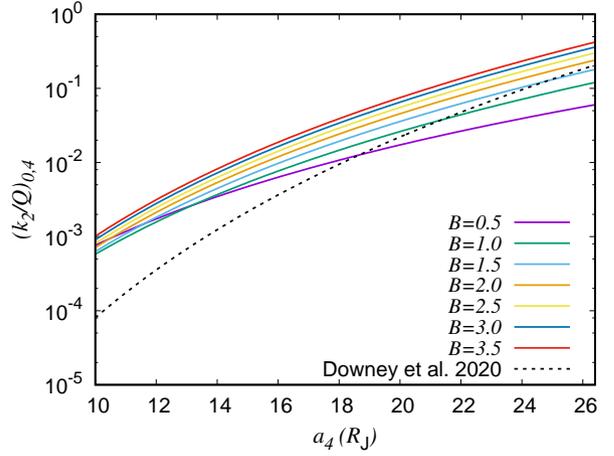}
   \caption{Value of Jupiter's effective dissipative parameter at the frequency of Callisto $(k_2/Q)_{0,4}$ as a function of Callisto's semi-major axis $a_4$, assuming that it is resonantly locked. The current value of $a_4$ is about $26.4$ $R_{\text{J}}$. Coloured curves are obtained using Eq.~\eqref{eq:fultheo} with different values of the parameter $B$. The black dashed curve shows the change in value of $(k_2/Q)_{0,4}$ considered in \citet{DOWNEY-etal_2020} for their simulations.}
   \label{fig:k2qcal}
\end{figure}

Since Io is the Galilean satellite closest to Jupiter, it experiences the largest dissipation among all four moons, so that tidal effects between Io and Jupiter have been thought to drive the orbital evolution of the whole moon system \citep{YODER-PEALE_1981,MALHOTRA_1991,LAINEY-etal_2009,LARI-etal_2020}. Through this mechanism, the momentum transferred from the spin of Jupiter to the orbit of Io and the energy loss due to the tidal friction within Io are redistributed to Europa and Ganymede via the Laplace resonance. However, the resonance locking theory presented by \citet{FULLER-etal_2016} proposes instead that the dissipation of Jupiter at the frequency of Callisto is very high (with today's effective quality factor $Q$ close to $1$), so that Callisto would be actually the satellite with the highest migration rate of the system. In the tidal theory of \citet{FULLER-etal_2016}, the orbital motion of the satellite is resonantly locked with an interior oscillation mode of the planet, and its migration timescale follows the planet's internal structure evolution.

Assuming that Callisto is resonantly locked with Jupiter, its semi-major axis over time $t$ follows a power law of the type (see \citealp{LAINEY-etal_2020})
\begin{equation}\label{eq:plaw}
   a_i(t)=a_0\left(\frac{t}{t_0}\right)^B,
\end{equation}
where $a_0$ is the current value of the moon's semi-major axis, $t_0\approx 4.57$ Gyrs is the age of Jupiter and $B=t_0/t_i^{\text{tide}}$. If the moon is resonantly locked, then its tidal timescale $t_i^{\text{tide}}$ is expected to evolve over time as the planet grows older, such that $B$ remains constant

While in classic theories dissipative parameters are considered constant \citep{MACDONALD_1964} or linear functions of the orbital frequency \citep{MIGNARD_1979}, for the resonance locking theory we have an effective value that depends strongly on the moon's semi-major axis \citep{FULLER-etal_2016,LAINEY-etal_2020}:
\begin{equation}\label{eq:fultheo}
   \left(\frac{k_2}{Q}\right)_{0,i}=\frac{B}{3}\frac{\sqrt{\mathcal{G}m_0}}{\mathcal{G}m_i}\frac{1}{R_\text{J}^5}\frac{1}{t_0}\left(\frac{a_i}{a_0}\right)^{1/B}a^{13/2}\,.
\end{equation}
Therefore, as shown in Fig.~\ref{fig:k2qcal}, the dissipative parameter $(k_2/Q)_{0,i}$ changes by many orders of magnitude during the moon's outward migration, growing from small values, when the moon is close to the planet, to large values, when the moon is farther away. Equation~\eqref{eq:fultheo} gives the instantaneous value of $(k_2/Q)_{0,i}$ that results in the migration law in Eq.~\eqref{eq:plaw} for the $i$th satellite. These formulas hold for a satellite that is not involved in any MMR, otherwise the effective value of $(k_2/Q)_{0,i}$ must be enhanced in order to compensate for the momentum transferred to the other resonant moons:
\begin{equation}\label{eq:fultheocapfac}
      \left(\frac{k_2}{Q}\right)^{\text{MMR}}_{0,i}=\left(\frac{k_2}{Q}\right)_{0,i}(1+f)\,.
\end{equation}
The enhancing factor $f$ depends on the satellites involved in the resonance and on the kind of resonant link. For simplicity, and since the tidal history of Callisto is not known yet, we define empirically $f$ so that $a_i$ in our simulations still follows the evolution described in Eq.~\eqref{eq:plaw}.

\begin{table}
\caption{Possible past initial conditions of the Galilean satellites that verify the Laplace resonance (semi-major axes are given in $R_\text{J}$ units and angles are in radians). The eccentricities of the resonant satellites are set to their forced value, and Callisto is placed close the 2:1 resonance with Ganymede (see text). The symbol * indicates that the variable is independent and can be chosen randomly.}
{\small
\begin{center}
\noindent\begin{tabular}{r|rrrr}
\hline
elements      & I & II & III & IV \\
\hline
$a_1$         &  4.0000 &  4.5000 &  5.0000 &  5.5000  \\
$a_2$         &  6.3596 &  7.1576 &  7.9554 &  8.7528  \\
$a_3$         & 10.1447 & 11.4232 & 12.7008 & 13.9775  \\
$a_4$         & 15.6073 & 17.5742 & 19.5396 & 21.5039  \\
$e_1$         &  0.0034 &  0.0034 &  0.0034 &  0.0034  \\
$e_2$         &  0.0085 &  0.0081 &  0.0078 &  0.0075  \\
$e_3$         &  0.0015 &  0.0015 &  0.0015 &  0.0015  \\
$e_4$         &  0.0050 &  0.0050 &  0.0050 &  0.0050  \\
$\varpi_1$         & 0.7980 & 5.5297 & 5.2698 & 2.6530 \\
$\varpi_2$         & 3.9416 & 2.3913 & 2.1251 & 5.7986 \\
$\varpi_3$         & 0.7128 & 5.1816 & 5.6236 & 2.4659 \\
$\varpi_4$         &      * &      * &      * &      * \\
$\lambda_1-2\lambda_2$ & 5.4844 & 0.7526 & 1.0139 & 3.6293 \\
$\lambda_2-2\lambda_3$ & 2.3420 & 3.8934 & 4.1555 & 0.4873 \\
$\lambda_3-2\lambda_4$ &      * &      * &      * &     * \\
\hline
  \end{tabular}
\end{center}
}
\label{tab:incond}
\end{table}

In order to compute the value of the parameter $B$, the tidal timescale of Callisto would be needed. Unfortunately, the migration rate of Callisto has not been measured yet, but only hypothesised by \citet{FULLER-etal_2016}. Moreover, there is also the possibility that Callisto is not currently resonantly locked with Jupiter, but that it was locked in the past. For these reasons, we explore different values of $B$ consistent with the tidal-locking scenario of \citet{FULLER-etal_2016}. We consider values of $B$ between $0.5$ and $3.5$, which would correspond to current values of $(k_2/Q)_{0,4}$ between $0.06$ and $0.42$ (see Fig.~\ref{fig:k2qcal}). Since from Juno mission data the tidal Love number of Jupiter is well known \citep{DURANTE-etal_2020,WAHL-etal_2020}, such an interval corresponds to $Q_{0,4}$ between about $10$ and $1$. These values are clearly out of the range given by the classic tidal theory \citep{GOLDREICH-SOTER_1966}, while they are in line with the resonance locking expectations. Indeed, \citet{FULLER-etal_2016} predicted a current value of the effective quality factor of Jupiter at the frequency of Callisto of the order of unity. Building on this prediction, \citet{DOWNEY-etal_2020} used a value around $3$ (i.e. $k_2/Q=0.2$) for their computations. These values imply that Callisto should currently be migrating outwards (or should have migrated in the past) at a rate of several tens of centimetres per year, that is even faster than Titan \citep{LAINEY-etal_2020}. If Callisto is still resonantly locked today, then it should have crossed the 2:1 resonant region with Ganymede less than $1$ Gyr ago \citep{DOWNEY-etal_2020}. As it is relatively a short time with respect to the lifetime of the satellites, it is reasonable to assume that the Laplace resonance between the three inner moons was already established during this event.

\subsection{Setup}
\label{subsec:set}

In order to explore the past dynamical evolution of the moons in the context of a fast migration of Callisto, we need to consider the possible past configurations of the four moons. As Io, Europa, and Ganymede also migrate over time (though much slower than what we assume for Callisto), we take initial conditions in such a way that their orbits are closer to Jupiter than they are today, and that they are trapped in the Laplace resonance, i.e. the three relations in Eq.~\eqref{lap2bod} are fulfilled. In Table~\ref{tab:incond}, we report some examples taking $a_1$ from $4.0$ to $5.5$ $R_{\text J}$ (numbered I, II, III and IV). These initial conditions represent possible states of the moons about $1$ to a few Gyrs in the past. Moreover, we place Callisto just below the 2:1 resonant region with Ganymede, at $a_3/a_4=0.65$. Given the fast migration of Callisto, $a_3/a_4$ decreases, pushing Callisto toward the 2:1 resonance (nominally at $a_3/a_4\approx 0.63$). While the orbital elements of the three inner moons are forced by the Laplace resonance, we have a certain freedom for the elements of Callisto. We set Callisto's initial eccentricity to $0.005$, which is slightly smaller than today's value; large values are not likely, because of the damping due to tidal friction. Finally, as the resonant encounter introduces chaos in the motion of the satellites \citep{LARI-etal_2020}, we sample the mean longitude of Callisto in the whole interval $[0:2\pi)$. Indeed, the slightest change in one of the model parameters would result in extreme variations of $\lambda_4$ after a few millions of years. The initial orbital phase of Callisto with respect to the three inner moons is therefore a completely free parameter. By sampling $\lambda_4$ in $[0,2\pi)$, we aim to build a complete picture of the diversity of outcomes that can be produced by the resonant crossing.

For what concerns the dissipative parameters, we set $(k_2/Q)_{0,i}$ $(i=1,2,3)$ and $(k_2/Q)_{1}$ to the values measured by \citet{LAINEY-etal_2009}, while for $(k_2/Q)_{i}$ $(i=2,3,4)$, we consider cases with small ($0.001$), medium ($0.005$) and large ($0.010$) tidal friction within the moons. Finally, as described in Sect.~\ref{subsec:tid}, we make $(k_2/Q)_{0,4}$ evolve following Eq.~\eqref{eq:fultheo}, considering values of $B$ between $0.5$ and $3.5$. All other dissipative parameters are assumed to remain constant during the timespan of our simulations (constant-$Q$ model).

Because of the large timespans involved (from hundreds of millions to billions of years) and the large separation of timescales between conservative dynamics and tidal dissipation effects, for such studies a tidal acceleration factor $\alpha$ is commonly employed \citep{TITTEMORE-WISDOM_1988,MALHOTRA_1991,LARI-etal_2020,CELLETTI-etal_2022}. This factor multiplies $c_i$, defined in Eq.~\eqref{eq:cmalho}, and allows to speed up the propagation. In the case of the future evolution of the Galilean satellites, with Ganymede migrating around $10$ cm/yr and Callisto steady, \citet{LARI-etal_2020} showed that using a factor $\alpha=10^2$ or $\alpha=10^3$ does not change significantly the statistics of the resonant encounter. With about the same migration rates, \citet{MALHOTRA_1991} and \citet{SHOWMAN-MALHOTRA_1997} used a factor $\alpha=10^3$ to explore the past evolution of the Galilean satellites. As in our nominal scenario Callisto migrates faster (several tens of centimetres per year), we use a conservative value and set $\alpha=10^2$, both in the averaged model and the $N$-body numerical integrations. In this way, we are inside the range of the absolute migration rate that \citet{LARI-etal_2020} explored in their computations and we are assured that the dynamics is not artificially altered by the acceleration factor. For the sake of clarity, we will present the results of the simulations in terms of the real physical time $t$, which is related to the integration time-variable $\tilde{t}$ through $t\approx\alpha\,\tilde{t}$.

\section{Outcomes of the resonant encounter}
\label{sec:res}

In this section, we show the evolution of the satellites as Callisto crosses the 2:1 MMR with Ganymede. All simulations and results presented in the section are obtained running the averaged model. As described by \citet{LARI-etal_2020}, because of the already existing resonances between the inner satellites, the 2:1 resonant region is surrounded by pure three-body MMRs which make the dynamics chaotic and non-trivial. Indeed, many outcomes are possible. As Callisto's orbit is diverging, the classic theory of two-body MMRs would predict that a capture is not possible (see e.g. \citealp{MURRAY-DERMOTT_2000}); however, the Galilean satellites dynamics is much more complicated than a mere succession of isolated two-body MMRs, and we see below that the actual dynamics defy this naive expectation.

Table~\ref{tab:stat} shows the statistics of outcomes for three different values of the moons' dissipative parameters. For each of these three experiments, $B$ is set to $1.5$ and $\lambda_4$ is sampled to $100$ equidistant values in $[0:2\pi)$. We classify the outcomes of the simulations in three main cases. In case A, Callisto is captured into resonance (two-body or pure three-body MMR) and the Laplace angle continues to librate. In case B, Callisto is captured into a pure three-body resonance and the Laplace angle stops librating and starts to circulate. In case C, Callisto crosses the resonance without being captured. It is worth noting that the classification used in the paper is similar to the one presented by \citet{LARI-etal_2020}, apart from case C, which did not appear in their study.

For each case (A, B, C) we will define below two subcases. The total percentage of a case X in Table~\ref{tab:stat} is therefore the sum of the percentages of X.1 and X.2. From these numbers, we can appreciate how the level of energy dissipation within the satellites affects the statistics of the resonant encounter: in particular, the non-capture of Callisto (case C) shows a greater probability when the energy dissipation within the moons is small, and the capture of Callisto together with the circulation of the Laplace angle (case B) shows a greater probability when the energy dissipation is large. Below we will refer to the average percentages.

\begin{figure}
   \centering
   \includegraphics[scale=0.43]{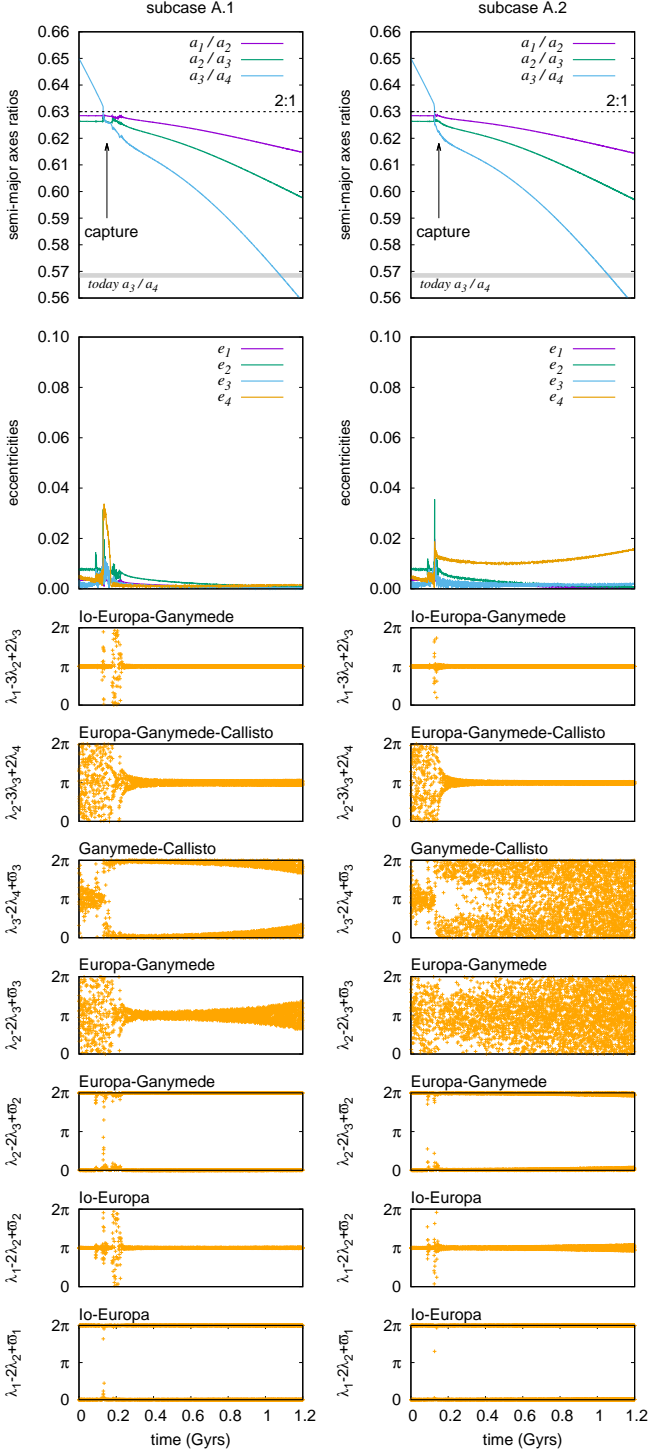}
   \caption{Examples of orbital evolutions for case A: Callisto is captured into resonance and the Laplace angle continues to librate. The left column shows a simulation where Callisto triggers a 2:1 MMR with Ganymede (subcase A.1). The right column shows a simulation where Callisto triggers a pure three-body MMR (Eq.~\ref{eq:3resa}) with Europa and Ganymede (subcase A.2). In both simulations, we set $B=1.5$ and $(k_2/Q)_i=0.005$ $(i=2,3,4)$.}
   \label{fig:caseA}
\end{figure}

\begin{figure}
   \centering
   \includegraphics[scale=0.43]{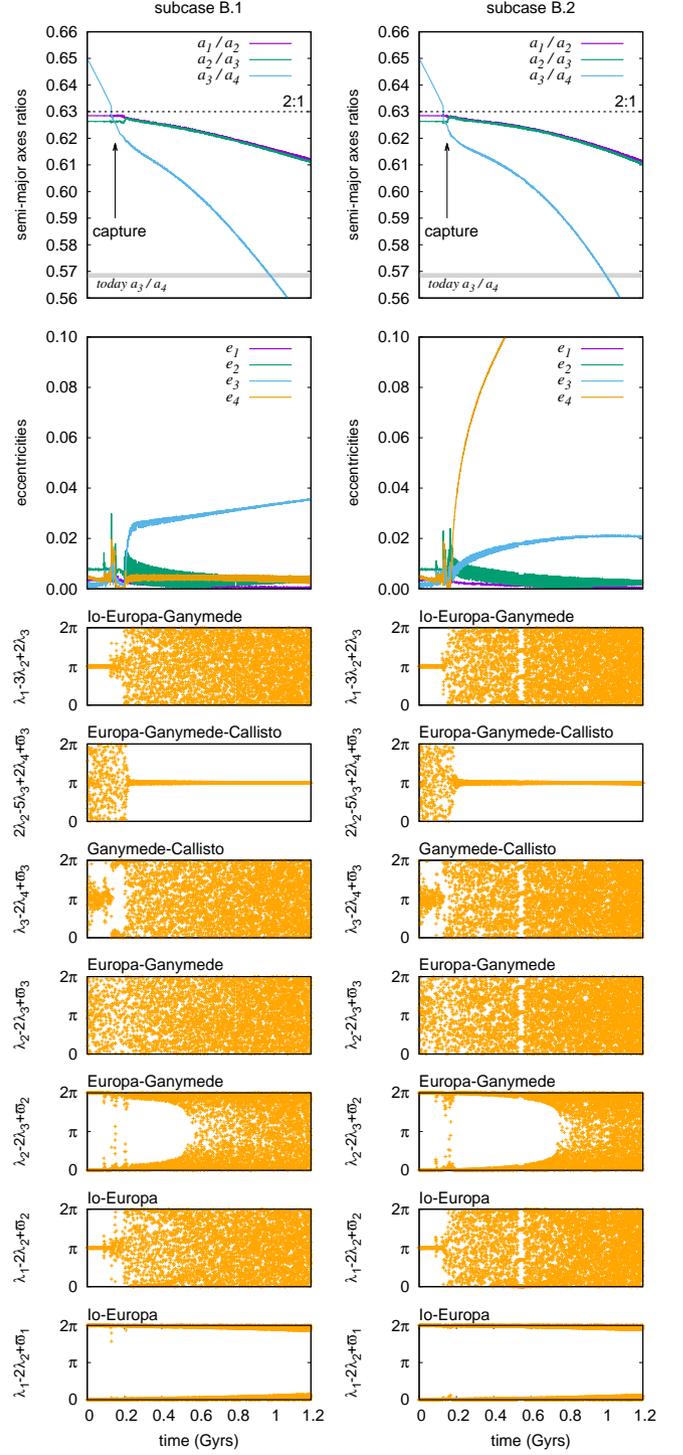}
   \caption{Examples of orbital evolutions for case B: Callisto is captured into resonance and the Laplace angle starts to circulate. The left column shows a simulation where the main increase in eccentricity is for Ganymede (subcase B.1). The right column shows a simulation where the main increase in eccentricity is for Callisto (subcase B.2). In both simulations, we set $B=1.5$ and $(k_2/Q)_i=0.005$ $(i=2,3,4)$.}
   \label{fig:caseB}
\end{figure}

\begin{figure}
   \centering
   \includegraphics[scale=0.43]{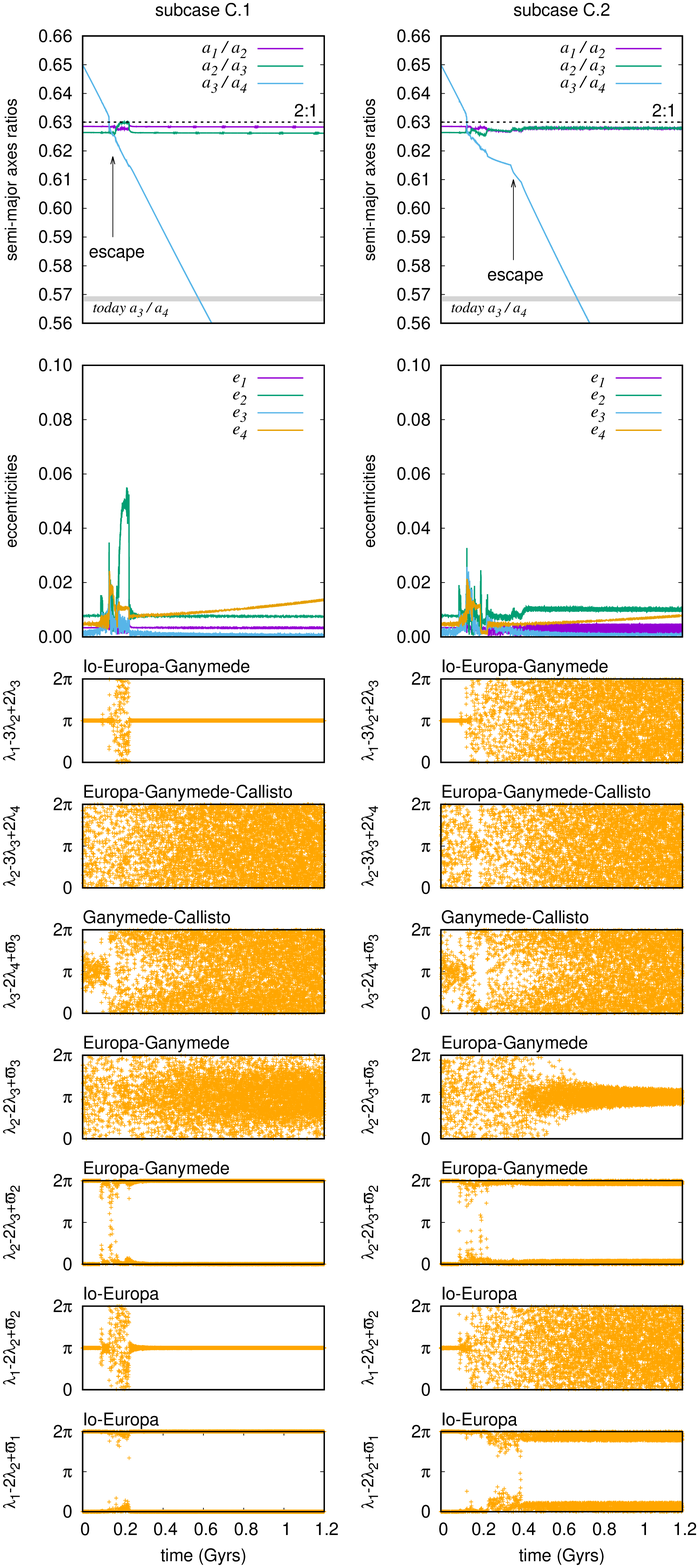}
   \caption{Orbital evolutions for case C: Callisto is not captured into resonance. The left column shows a simulation where the Laplace resonance is preserved (subcase C.1). The right column shows a simulation where the Laplace resonance is disrupted (subcase C.2). In both simulations, we set $B=1.5$ and $(k_2/Q)_i=0.001$ $(i=2,3,4)$.}
   \label{fig:caseC}
\end{figure}

\begin{figure}
   \centering
   \includegraphics[scale=0.7]{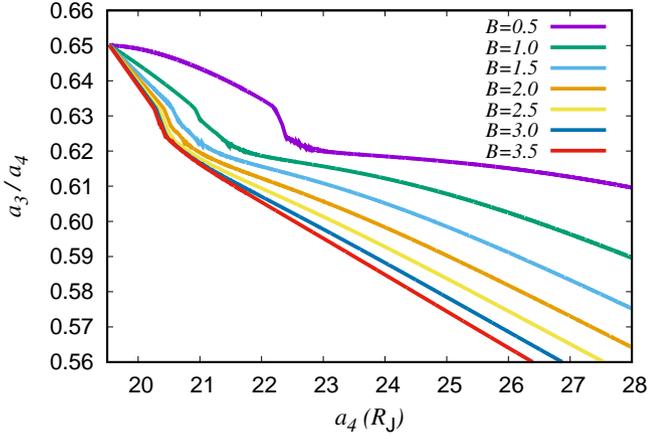}
   \caption{Evolution of the semi-major axes ratio $a_3/a_4$ in function of $a_4$ for outcome A, starting from initial conditions III of Table~\ref{tab:incond} and using different values of the tidal parameter $B$. The system evolution is faster when $B$ is larger: the evolution shown takes about $0.3$ Gyrs for $B=3.5$ and $2.0$ Gyrs for $B=0.5$.}
   \label{fig:diffB}
\end{figure}

\begin{figure}
   \centering
   \includegraphics[scale=0.75]{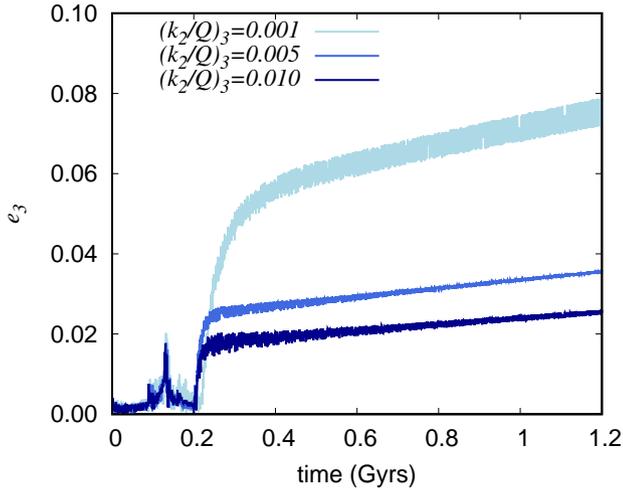}
   \caption{Evolution of Ganymede's eccentricity for simulations of subcase B.1, using different values of the dissipative parameter $(k_2/Q)_3$. Low dissipation within the moon allows a larger increase of its eccentricity.}
   \label{fig:eccga}
\end{figure}

In Fig.~\ref{fig:caseA},~\ref{fig:caseB} and~\ref{fig:caseC}, we report the typical evolution of the semi-major axes ratios, eccentricities and resonant angles for cases A, B and C, respectively. As shown in Fig.~\ref{fig:diffB} and Fig.~\ref{fig:eccga} different values of $B$ mainly affect the variation rate of the semi-major axes (which sets the global timescale of the system evolution), while parameters $(k_2/Q)_i$ mainly affect the eccentricities. This is an expected result: the equilibrium values of the moons' eccentricities result from a balance between the effects of the MMRs (that force the eccentricity to not be zero) and the level of eccentricity damping (that tends to decrease the eccentricity to zero).

\begin{table}
\caption{Statistics of the outcomes of the resonant encounter with Callisto moving on a divergent orbit; for this analysis, we set $B=1.5$ and we considered different values for the dissipative parameters $(k_2/Q)_i$ $(i=2,3,4)$. Different cases are defined as follows: case A, Callisto is captured into resonance and the Laplace angle continues to librate; case B, Callisto is captured into resonance and the Laplace angle stops librating; case C, Callisto in not captured into resonance (we refer to the main text for the definition of the subcases). Boldface characters show the outcomes that match the current orbital configuration of the moons.}
{\small
\begin{center}
\noindent\begin{tabular}{c|r|r|r|r}
\hline
$(k_2/Q)_{2,3,4}$  & $0.001$ & $0.005$ & $0.010$ & average \\
\hline
A.1 & $29\%$ & $48\%$ & $70\%$ & $49\%$ \\
A.2 & $55\%$ & $31\%$ &  $3\%$ & $29\%$ \\
B.1 &  $5\%$ & $10\%$ & $11\%$ &  $9\%$ \\
B.2 &  $0\%$ &  $5\%$ & $12\%$ &  $6\%$ \\
\bf{C.1} & $\mathbf{5\%}$ & $\mathbf{1\%}$ & $\mathbf{0\%}$ & $\mathbf{2\%}$ \\
C.2 &  $6\%$ &  $5\%$ &  $4\%$ &  $5\%$ \\
\hline
  \end{tabular}
\end{center}
}
\label{tab:stat}
\end{table}

From Table~\ref{tab:stat}, we see that, in most simulations ($93\%$, case A plus case B), Callisto is captured into resonance with the other satellites, forming an 8:4:2:1 resonant chain. From Fig.~\ref{fig:caseA} and Fig.~\ref{fig:caseB}, we can note how a new resonant angle involving Callisto's mean longitude starts to librate. The slope of $a_3/a_4$ changes once it reaches about $0.63$: the reason is that, once Callisto enters into resonance, the huge amount of angular momentum that it gains from Jupiter is redistributed also to the other three moons. As a result, the migration of Callisto slows down, while the migration rate of the three inner moons increases. Moreover, the evolution of the semi-major axes greatly differs from what was found by \citet{LARI-etal_2020} in the case of convergent orbits (see Fig.~\ref{fig:evolconv}): in their analysis, $a_3/a_4$ converged to the resonant value ($\approx 0.63$) and remained fixed after the capture of Callisto. Here, instead, $a_3/a_4$ decreases as Callisto continues its fast outward migration and it moves away from the nominal value of the two-body 2:1 MMR. As shown in Figs.~\ref{fig:caseA} and~\ref{fig:caseB}, the other ratios $a_1/a_2$ and $a_2/a_3$ also decrease over time. This is typical of three-body MMRs, which allow for variations of semi-major axis ratios even though all bodies still remain locked in resonance.

Cases A and B differ by the behaviour of the Laplace angle $\lambda_1-3\lambda_2+2\lambda_3$. As shown in Fig.~\ref{fig:caseA}, case A also features a new Laplace-like resonance between Europa, Ganymede and Callisto:
\begin{equation}
   \label{eq:3resa}
   \lambda_2-3\lambda_3+2\lambda_4\sim \pi,
\end{equation}
which is the same relation as the current Laplace resonance between Io, Europa and Ganymede. Equation~\eqref{eq:3resa} can be either the result of the sum of two-body resonances triggered by the couples Europa-Ganymede and Ganymede-Callisto (subcase A.1), or it can indicate a pure three-body resonance between the three outer moons (subcase A.2). After the resonant encounter, the eccentricity of Callisto can jump to different values: for simulations of kind A.1, the eccentricity is forced to nearly $0$ because of the two-body resonant link; while for simulations of kind A.2, it is excited to higher values, and then it slowly increases or decreases, depending on which term between $(k_2/Q)_{(0,4)}$ and $(k_2/Q)_4$ is dominant (see Eq.~\eqref{eq:disse}). In both cases, the eccentricities of the inner moons decrease to values lower than their current values forced by the Laplace resonance. This is due to the decrease in the ratios $a_1/a_2$ and $a_2/a_3$, which makes the regression rates of the moons' pericentres increase to preserve the resonances, and they are inversely proportional to the forced eccentricities (see e.g. \citealp{SINCLAIR_1975}). Finally, from Fig.~\ref{fig:caseA}, we can see that the initial two-body resonant angles $\lambda_1-2\lambda_2+\varpi_1$, $\lambda_1-2\lambda_2+\varpi_2$ and $\lambda_2-2\lambda_3+\varpi_2$ continue to librate, even though the libration amplitudes increase over time.

In simulations of case B, apart from the circulation of the Laplace angle, Callisto is captured into a pure three-body MMR with Europa and Ganymede which is different from the relation in Eq.~\eqref{eq:3resa} (see Fig.~\ref{fig:caseB}). The most common pure three-body MMR obtained in our simulations is
\begin{equation}
   \label{eq:3resb}
   2\lambda_2-5\lambda_3+2\lambda_4+\varpi_3\sim \pi,
\end{equation}
which was also the most likely pure three-body resonance found by \citet{LARI-etal_2020} in their study of the future evolution of the Galilean satellites. For this resonance, the eccentricities of the outer moons can increase greatly. The actual eccentricity value reached depends on the values of the moons' dissipative parameters $(k_2/Q)_i$. For some simulations (subcase B.1, see Fig.~\ref{fig:caseB}), the main effect is on the eccentricity of Ganymede. As shown in Fig.~\ref{fig:eccga}, the eccentricity of Ganymede reaches values between $0.02$ and $0.08$, depending on the magnitude of the dissipative parameters. For case B.2, on the contrary, the main effect is on the eccentricity of Callisto, which can rapidly grow over $0.1$ and then it pumps the eccentricity of Ganymede up to around $0.02$. As already described by \citet{LARI-etal_2020}, the different behaviour of the eccentricities for the two subcases is related to the combinations $\varpi_2-\varpi_3$ and $\varpi_3-\varpi_4$, which can librate around $0$ or $\pi$. Depending on the libration of one or both these combinations, several three-body resonances can be activated and add up to the one in Eq.~\eqref{eq:3resb}. Finally, only the two-body resonance between Io and Europa holds on ($\lambda_1-2\lambda_2+\varpi_1\sim 0$), while the resonant angles $\lambda_1-2\lambda_2+\varpi_2$ and $\lambda_2-2\lambda_3+\varpi_2$ start to circulate, leading to the disruption of the Laplace resonance as we know it today (see Fig.~\ref{fig:caseB}).

Surprisingly, in both cases A and B, the resonances involving Callisto survive for a very long time, even though $a_3/a_4$ moves quite far from the nominal value of the nominal two-body 2:1 MMR. Only for very low values ($a_3/a_4<0.5$, not shown), the resonance completely disappears and the four-body resonant chain breaks down. This means that Callisto can spend from hundreds of millions to billions of years in resonance with the other three Galilean moons, even if it is moving on a divergent orbit.

In the rest of the simulations (only $7\%$, case C), Callisto is not captured into resonance. From Fig.~\ref{fig:caseC}, we can appreciate how $a_3/a_4$ recovers its initial slope after the crossing, which implies that the satellite does not remain locked in any resonant link with the other moons. In this situation, Callisto does not redistribute the angular momentum that it gains from Jupiter, and so $a_3/a_4$ reaches today's value much faster than in cases A or B. Also for these simulations, we observe two subcases: in the first one (subcase C.1), Callisto crosses the 2:1 resonant region and the Laplace resonance is preserved; in the second one (subcase C.2), it crosses the 2:1 resonant region and the Laplace resonance is disrupted. In the first subcase, apart from the libration of the angles involved in the Laplace resonance, we can note how after a relatively brief period of time where the moons' eccentricities can be excited (even to quite high values), they settle again to the values forced by the resonances.

Therefore, we actually found a class of simulations that matches qualitatively the current configuration of the Galilean satellites. Indeed, subcase C.1 is the scenario expected by \citet{DOWNEY-etal_2020}, but, although it is the most straightforward one, we obtained just a total of $2\%$ simulations of this kind. All the other simulations do not reproduce directly today's configuration of the moons. In cases A and B, Callisto remains locked into resonance, which evidently is in contrast with the current situation; while, in case C.2, Callisto is not captured and the Laplace resonance is in part disrupted. As the capture of Callisto is the most likely pathway, in Sect.~\ref{sec:pat}, we investigate whether some additional dynamical mechanisms, that begin to play a role as the system departs from the exact 2:1 commensurabilities, could help us to restore the configuration of the system after the capture of Callisto.

\section{Evolution through a 4-body resonant chain}
\label{sec:pat}

In Sect.~\ref{sec:res}, we found that if Callisto migrates faster than the other moons and crossed the 2:1 MMR with Ganymede in the past, the most likely outcome is the capture into a four-body resonant chain with the other satellites (cases A and B). Differently from the convergent evolution presented by \citet{LARI-etal_2020} and showed in Fig.~\ref{fig:evolconv}, after the capture the ratio $a_3/a_4$ decreases, passing from $0.63$ to $0.56$ in several hundreds of millions of years (or up to a few billions of years, depending on the exact value of $B$; see Fig.~\ref{fig:diffB}). Since all semi-major axes ratios move away from the nominal 2:1 MMRs, we expect that our averaged model gradually becomes less accurate after the capture of Callisto, as it is designed to be valid in the vicinity of the 2:1 MMR chain and lacks other potential resonant terms.

\begin{figure}
   \centering
   \includegraphics[scale=0.65]{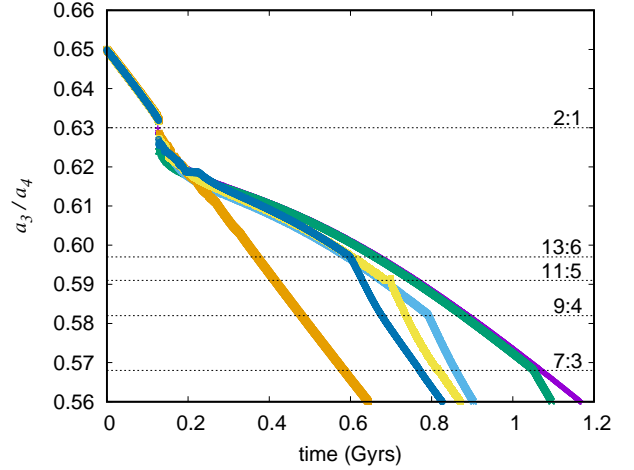}
   \caption{Examples of Callisto's capture into and escape from the four-body resonant chain. Each curve has been obtained from an un-averaged N-body integration with parameter $B=1.5$. The colours correspond to six different values of $(k_2/Q)_i$ $(i=1,2,3)$. High-order MMRs between Callisto and Ganymede are labelled.}
   \label{fig:exitres}
\end{figure}

For this reason, we explore the subsequent evolution of the Galilean satellites after the capture of Callisto by running un-averaged $N$-body numerical integrations. These simulations require a much larger computation time than the ones obtained with the averaged model, therefore their number is quite limited.

For comparison purpose, and to avoid the delicate task of converting back averaged orbital elements to un-averaged coordinates, we begin our $N$-body numerical integrations before the capture of Callisto. Our $N$-body simulations closely match the results obtained in Sect.~\ref{sec:res} during the stage of Callisto's capture; this means that our averaged model did capture the essence of the dynamics in a neighbourhood of the nominal 2:1 two-body MMRs. However, as $a_3/a_4$ continues to decrease substantially, new dynamical features show up. Indeed, as shown in Fig.~\ref{fig:exitres} by the abrupt change of the slope of $a_3/a_4$, at certain points of the evolution the resonant link of Callisto can break down.

These points correspond to the crossing of high-order ($>3$) MMRs between Ganymede and Callisto, which can destabilise the system and release Callisto from the resonance. From that point on, the migration of Callisto does not push outwards the other moons anymore, as the angular momentum that it gains from Jupiter is no longer exchanged with the other moons. In Fig.~\ref{fig:exitres}, we show the nominal position of the main high-order MMRs that Callisto crosses.

\begin{figure}
   \centering
   \includegraphics[scale=0.75]{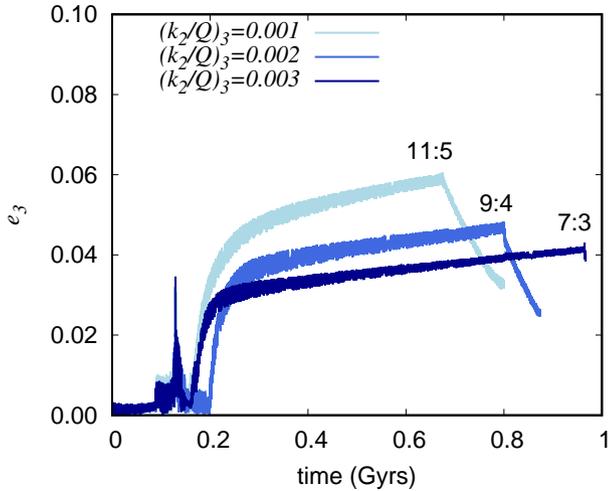}
   \caption{Evolution of Ganymede's eccentricity for simulations of subcase B.1, obtained with un-averaged $N$-body numerical integrations. For each curve, the abrupt change corresponds to the escape of Callisto from the four-body resonance chain.}
   \label{fig:eccganum}
\end{figure}

\begin{figure}
   \centering
   \includegraphics[scale=0.75]{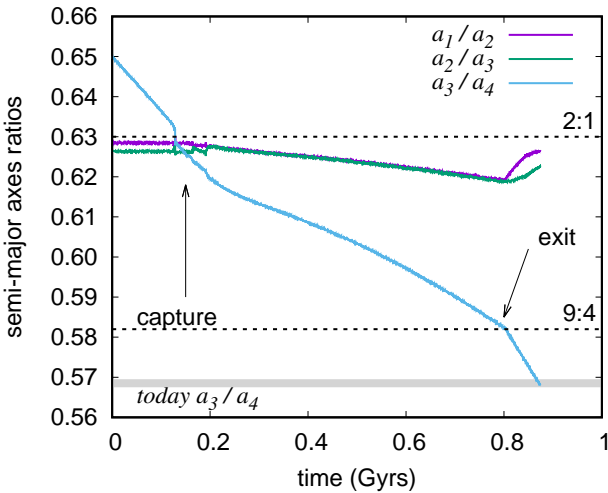}
   \caption{Simulation of the kind B.1, in which the four-body resonant chain breaks down because of the crossing of the 9:4 MMR between Ganymede and Callisto. After Callisto's escape, $a_1/a_2$ and $a_2/a_3$ converge towards the 2:1 MMR nominal value, but do not have time to reach it before $a_3/a_4$ crosses its current value.}
   \label{fig:casestop}
\end{figure}

As shown in Fig.~\ref{fig:eccganum}, the exact resonance that produces the escape of Callisto depends on the values of the moons' eccentricities, as they set the strength of these high-order MMRs. As explained in Sect.~\ref{sec:res}, the eccentricities reached by the satellites depend on their internal tidal dissipation. For high values of the eccentricities (mainly case B), we find that resonances 13:6 and 11:5 have a good probability to break the resonant chain. For smaller values of eccentricities (mainly case A), only stronger resonances, like 7:3 and 5:2, manage to release Callisto. As the current configuration of the system is close to the 7:3 resonance between Ganymede and Callisto, we can envision two distinct pathways.

In the first pathway, Callisto escapes the resonant chain early, through a MMR of order higher than 7:3 (for instance 9:4, 11:5, or 13:6). In this case, the migration rate of Callisto must remain high after its escape, because we still need $a_3/a_4$ to decrease down to its current value. However, Callisto's escape from the resonant chain does not guarantee that the current configuration of the system will be reproduced. Although Callisto is not in resonance anymore, we also need the eccentricities of the outer satellites to be damped to their current low values, and the Laplace resonance must be restored in case it was disrupted (case B). Moreover, the ratios $a_1/a_2$ and $a_2/a_3$ must retrieve their equilibrium values close to the nominal 2:1 two-body MMRs. This last point arises naturally as a result of the moons' eccentricity damping; however, this recovery of the Laplace resonance must happen before $a_3/a_4$ reaches it current value. In the example shown in Fig.~\ref{fig:casestop}, the convergence of $a_1/a_2$ and $a_2/a_3$ towards the nominal 2:1 level does happen, but it is too slow compared to the decrease in $a_3/a_4$. In order to counteract this effect, we need Callisto to escape the resonant chain earlier, for instance through the resonances 11:5 or 13:6. In order to allow for such high-order MMRs, however, the eccentricity damping of Ganymede must be lower (see Fig.~\ref{fig:eccganum}), which in turn slows down the recovery of the Laplace resonance. Hence, reproducing all features of the current configuration of the moons is possible through this first pathway, but it requires some fine tuning, such as a timely variation in the dissipative parameters.

In the second pathway, Callisto escapes the resonant chain later on, through the 7:3 MMR or a lower-order resonance (for instance, 5:2 or 3:1). In this case, the migration rate of Callisto must strongly decrease after its escape, so that the three other moons can catch up as they migrate outwards; this would increase $a_3/a_4$ again up to its current value.

An abrupt change in the dissipative parameters is clearly an ad hoc assumption, but it is unavoidable here to reproduce the current configuration of the moons. From a physical point of view, a change in the energy dissipation is possible, and it has already been invoked in previous studies of the Galilean system (see \citealp{SHOWMAN-MALHOTRA_1997}). \citet{HUSSMANN-SPOHN_2004} showed that the coupling between thermal and orbital evolution of the moons can produce wide periodic oscillations in their dissipation rate. Moreover, large variations in the dissipative parameters are expected with the activation of dynamical processes within the celestial bodies. In the case of satellites, it can be due for instance to resonances in moon-moon tides \citep{HAY-etal_2020}. Moreover, an abrupt change in the tidal dissipation within the planet may also happen at some point, because of a release of Callisto from the tidal resonance locking mechanism of \citet{FULLER-etal_2016}.

\begin{figure}
   \centering
   \includegraphics[scale=0.43]{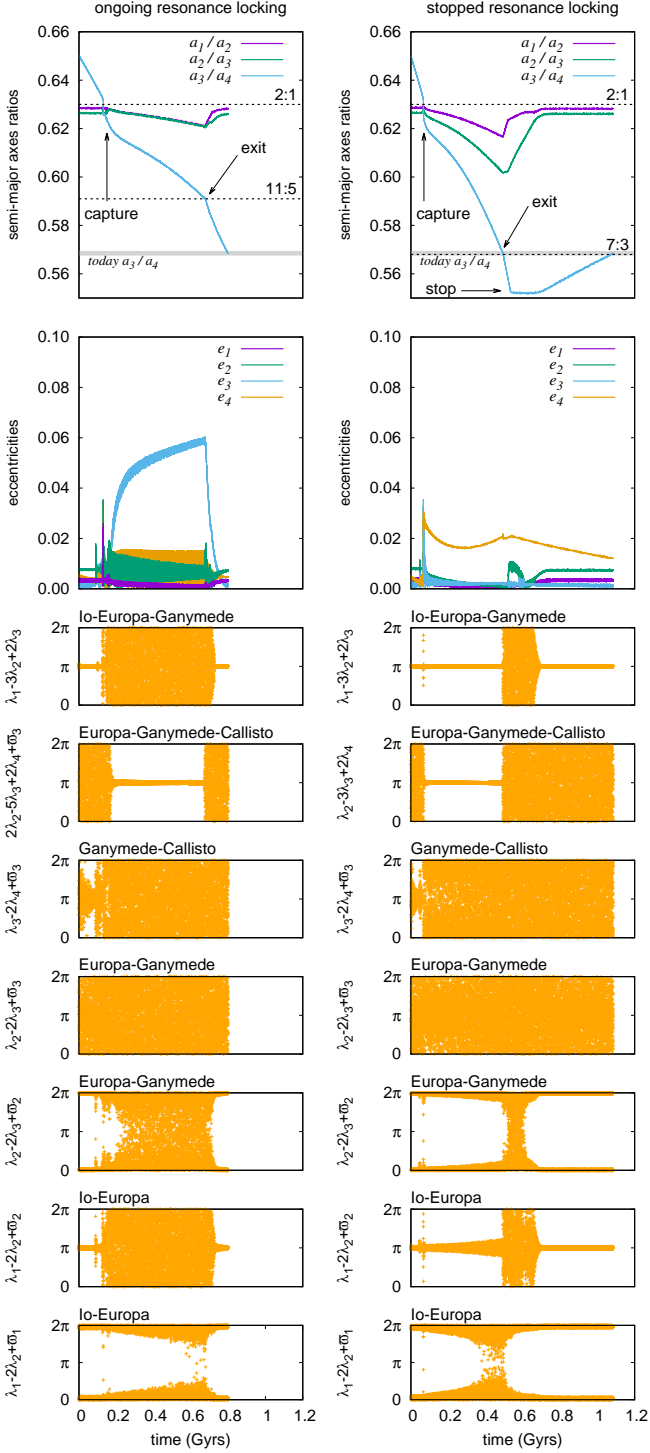}
   \caption{Orbital evolutions of the Galilean satellites, obtained with $N$-body numerical integrations, that match qualitatively the current configuration of the system. The left column shows a simulation where the moons follow an evolution of kind B.1 and then Callisto exits from the resonant chain through the crossing of the 11:5 MMR. The right column shows a simulation where the moons follow an evolution of kind A.2, then Callisto exits from the resonant chain through the crossing of the 7:3 MMR and we stop its fast outward migration. For the first simulation, we
set $B=1.5$ and $(k_2/Q)_3=0.001$ (then increased to $0.007$); for the second one, $B=3.1$ and $(k_2/Q)_4=0.008$.}
   \label{fig:casesdivnum}
\end{figure}

For comparison, Fig.~\ref{fig:casesdivnum} shows one successful simulation for each of the two possible dynamical pathways: on the left, Callisto escapes early from the resonant chain; on the right Callisto escapes later on, and its tidal migration is switched off after some time.

\begin{figure}
   \centering
   \includegraphics[scale=0.49]{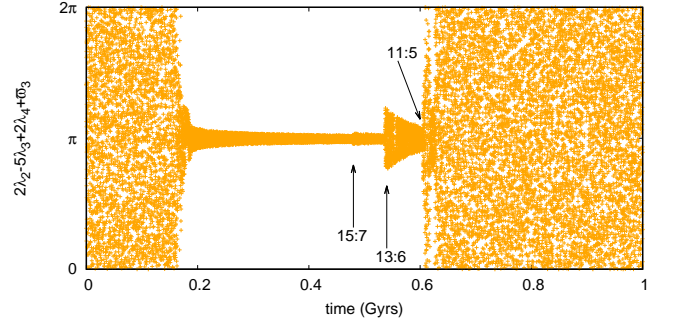}
   \caption{Evolution of the resonant angle $2\lambda_2-5\lambda_3+2\lambda_4+\varpi_3$ from the capture to the exit of Callisto, obtained with an un-averaged $N$-body numerical integration. The crossing of two-body MMRs between Gaymede and Callisto are labelled.}
   \label{fig:exit115}
\end{figure}

The first one is obtained starting from initial conditions III of Table~\ref{tab:incond}, initially setting $B=1.5$ and $(k_2/Q)_3=0.001$, and it corresponds to an evolution of the kind B.1: Callisto is captured into the three-body resonance given by $2\lambda_2-5\lambda_3+2\lambda_4+\varpi_3\sim \pi$ and the eccentricity of Ganymede increases up to large values (about $0.06$). The first part of the evolution is very similar to the one already presented in Fig.~\ref{fig:caseB}, but because of the crossing of high-order MMRs, the resonant angle involving Callisto is excited. A zoom-in view of this phenomenon is presented in Fig.~\ref{fig:exit115}. As $a_3/a_4$ decreases, the crossing of different two-body MMRs between Callisto and Ganymede has a direct noticeable effect on the amplitude of the three-body resonance angle. Each resonance produces a kick, which is then damped through tidal dissipation. The magnitude of the successive kicks increases as the order of the MMR encountered decreases, until a strong enough resonance (here, 11:5) ejects Callisto from the three-body MMR. Once Callisto exits from the resonance, the ratio $a_3/a_4$ decreases faster, while $a_1/a_2$ and $a_2/a_3$ move back to the 2:1 nominal resonant value, so that also the forced eccentricities of Io and Europa increase up to their current values. The eccentricity of Ganymede, which reached high values when Callisto was still in resonance, is also damped because of the tidal dissipation within the moon: in order to reach a value $e_3\approx 0.001$ in about $100$ Myrs, which is the time for $a_3/a_4$ to go from $0.591$ to $0.568$ (considering $B=1.5$), $(k_2/Q)_3$ must be at least $0.007$. Under this condition, when $a_3/a_4$ reaches today's value, we arrive very close to the current configuration of the system, with similar eccentricities and semi-major axes (see Table~\ref{tab:match}), the Laplace angle and all other angles in Eq.~\eqref{lap2bod} librate, and Callisto is not involved in any resonance anymore, as observed. If the Galilean system did follow this dynamical pathway, then it means that Callisto escaped recently from the resonant chain, and that all four moons reached their equilibrium configuration even more recently through tidal damping. This relatively recent increase and following damping of the eccentricities of the outer moons can also provide an explanation for the residual free eccentricities of Ganymede and Callisto \citep{SINCLAIR_1975,DOWNEY-etal_2020}.

A variant of this scenario can involve a large increase of Callisto's eccentricity (subcase B.2). In this case, the escape of Callisto from the resonant chain may happen earlier, as $e_4$ reaches values much larger than $e_3$, which strengthens high-order two-body MMRs. Simulations show that the 13:6 MMR can be strong enough for this. However, also in this case we would need to tune the dissipative parameters, and abruptly increase the eccentricity damping of Callisto in order for it to reach its current value in time. Moreover, a large increase in Callisto's eccentricity would result in a quite recent large tidal friction within the moon, which seems to contradict the almost complete lack of geological activity, as testified by its heavily cratered surface \citep{GREELEY-etal_2000}. For these reasons, we do not explore further evolutions involving a large increase in Callisto's eccentricity.

The second simulation in Fig.~\ref{fig:casesdivnum} gives an example of the second possible dynamical pathway. We obtained it starting from initial conditions II of Table~\ref{tab:incond}, setting $B=3.1$ and $(k_2/Q)_i=0.008$, and it corresponds to an evolution of the kind A.2: Callisto is captured into the three-body resonance given by $\lambda_2-3\lambda_3+2\lambda_4\sim \pi$ and the eccentricity of Callisto jumps to about $0.002$. As all eccentricities remain relatively small, high-order MMRs with Ganymede are not strong enough to kick Callisto out of the resonant chain. In this example, Callisto escapes only when the 7:3 MMR is crossed: in this case, $a_3/a_4$ is already close to today's value, and the system needs some time to resettle into today's configuration. Therefore, we continue the evolution until Callisto reaches its current semi-major axis (about $26.4$, and $a_3/a_4<0.568$) and we assume that the resonance lock of Callisto with the interior mode of Jupiter breaks down, so that its orbital expansion stops. Such a breaking is indeed expected when the moon is too far away from the planet and the amplitude of the mode required to sustain the resonance locking becomes too large \citep{FULLER-etal_2016}). Subsequently, the tidal dissipation between Io and Jupiter becomes the dominant driver of the evolution of the semi-major axis ratios. This makes $a_3/a_4$ increase, because the three inner moons migrate away faster than Callisto. Therefore, after a few hundreds of millions of years the Galilean moons reach orbits similar to their current ones (see Table~\ref{tab:match}). Differently from the previous dynamical pathway, both the escape of Callisto from the resonant chain and the recovery of the equilibrium configuration for the inner moons happen earlier, perhaps several hundreds of millions of years ago.

In Table~\ref{tab:match}, we report the final values of the semi-major axes and eccentricities for both simulations. They are clearly close to their observed mean values, but $e_1$ and $e_2$ are a bit smaller. The reason is that, if we assume the values the dissipative parameters measured by \citet{LAINEY-etal_2009} and the inward migration of Io, the eccentricities of the moons are not steady, but their values change because of the shift of the resonance centres. As shown in \citet{LARI-etal_2020}, after Io stops its inward migration and all moons evolve outwards (as in our simulations), the eccentricities settle to lower equilibrium values ($0.0034$ for Io and $0.0074$ for Europa), which match the ones in Table~\ref{tab:match}.

\begin{table}
\caption{Mean orbital elements of the Galilean satellites at the time $a_3/a_4=0.5685$ for the two simulations described in Sect.~\ref{sec:pat} and represented in Fig.~\ref{fig:casesdivnum}.}
{\small
\begin{center}
\noindent\begin{tabular}{r|rrrr}
\hline
element & Io & Europa & Ganymede & Callisto \\
\hline
1)\hspace{12pt} $a$  & $5.8565$ & $9.3215$ & $14.8925$ & $26.1945$ \\
$e$  & $0.0033$ & $0.0072$ & $0.0012$  & $0.0048$ \\
\hline
2)\hspace{12pt} $a$  & $5.9037$ & $9.3987$ & $15.0083$ & $26.4044$ \\
$e$  & $0.0033$ & $0.0074$ & $0.0012$  & $0.0120$ \\
\hline
  \end{tabular}
\end{center}
}
\label{tab:match}
\end{table}

\section{Conclusion}
\label{sec:con}

The tidal resonance locking theory of \citet{FULLER-etal_2016} and the fast migration of Titan measured by \citet{LAINEY-etal_2020} open new avenues in our understanding of the Solar System dynamics. A fast migration has also been proposed for Callisto around Jupiter \citep{FULLER-etal_2016}, but it was unclear whether this hypothesis could be in agreement or not with the current peculiar configuration of the Galilean satellite system. A preliminary study of the effects of a fast migration for Callisto has been conducted by \citet{DOWNEY-etal_2020}. They showed that if Callisto migrates faster than the other moons, then it should have crossed the 2:1 MMR with Ganymede in the past. Their results, however, are limited to order-of-magnitude estimates, and a statistical analysis of the possible outcomes of this chaotic event was still missing. Here, we explored the past orbital dynamics of the Galilean satellites through numerical integrations and classified the possible effects of Callisto's fast migration. As the resonance crossing event is expected to be quite recent, we have assumed that the Laplace resonance between the three inner moons was already established at that time. This is consistent with formation models predicting a primordial origin for the Laplace resonance (see \citealp{PEALE-LEE_2002}).

As the resonant encounter between Callisto and Ganymede is chaotic, we cannot know for sure which dynamical pathway the system may have followed, so we must conduct a statistical analysis of the possible outcomes. To this aim, we have developed an averaged model of the Galilean satellites which is valid in the vicinity of the 2:1 resonance. This allows us to compute hundreds of simulations in a reasonable amount of time. We found that Callisto can cross the resonance without being captured, and the Laplace resonance can be preserved during this event, with only temporary excitation of the moons' eccentricities. However, this outcome only represents a small fraction of the possible trajectories (about $2\%$).

In the vast majority of cases Callisto is captured in resonance despite its divergent migration. This occurs while Callisto passes through the jungle of three-body MMRs that surrounds the 2:1 commensurability with Ganymede. The capture of Callisto thus leads to a resonant chain that involves all four Galilean moons. From this point on, the fast migration of Callisto becomes the main driver of the moons' migration because the angular momentum gained by Callisto from Jupiter is redistributed among all four moons. This strong outward pull of Callisto, however, cannot go on forever, because it makes each pair of adjacent moons gradually drift away from the exact 2:1 commensurability. We explored this drift process through un-averaged numerical simulations. Our simulations show that while the moons remain trapped in three-body MMRs, Callisto then crosses a high-order two-body MMR with Ganymede (e.g. 7:3, 9:4, 11:5...). Depending on the eccentricity values of the moons at that time, this final resonance crossing can free Callisto from the resonant chain after several hundreds of millions of years. Hence, we may still retrieve a configuration that is qualitatively similar to the current Galilean moons even if Callisto is temporarily captured. Yet, our simulations show that in this case, it is challenging to retrieve altogether today's resonant relations, semi-major axis ratios, and eccentricities of the moons.

If the eccentricity of Ganymede grows to large values (which happens in $9\%$ of simulations, provided that the tidal dissipation within Ganymede is low enough), then Callisto escapes early from the resonant chain and it can migrate to its current location while the Laplace resonance relaxes to its observed state. In this situation, Callisto may still be migrating fast today. If instead, the eccentricities of the moons remain quite low (which happens in $78\%$ of simulations), then Callisto escapes late from the resonant chain, and only after having crossed the current ratio $a_3/a_4$. In this second situation, the fast migration of Callisto must necessarily have stopped at some point, so that the other three Galilean moons can catch up and restore the current ratio $a_3/a_4$. Even though these two situations are possible as a results of the chaotic resonant encounter, they require a fine tuning of the dissipative parameters of the moons in order for them to retrieve their current eccentricities by the time the ratio $a_3/a_4$ reaches its current value.

The temporary increase in the satellites' eccentricity may be a way to distinguish between these different pathways. The eccentricity of Ganymede, in particular, can grow up to $0.06$, which would result in a strong tidal friction within the satellite. This enhanced interior heat could be the source of tectonic resurfacing and the smooth bright terrains on the surface of Ganymede \citep{MALHOTRA_1991,SHOWMAN-MALHOTRA_1997,SCHUBERT-etal_2004}.

Our current knowledge of tidal dissipation in the Jovian system comes from astrometric observations of the moons, which allowed to estimate the strong dissipation within Io \citep{LAINEY-etal_2009}, that is the main source of its observed heat flow \citep{VEEDER-etal_1994}. However, because of the limitations of the available data sets, we do not have precise estimates of the dissipative parameters related to the other moons. In the near future, JUICE and Europa Clipper space missions will visit the Jovian system, performing multiple flybys of the Galilean satellites. They will provide important observations that will allow to investigate further the energy dissipation in the Jovian system \citep{DIRKX-etal_2017,LARI-MILANI_2019}. In particular, with the combination of precise radio-science from space missions and other data sets (like astrometric observations) which cover wider timespans, it could be possible to estimate the dissipative parameters of all four Galilean satellites and of Jupiter at the different orbital frequencies of the moons. In this way, it will be possible not only to confirm the strong tidal dissipation within Io, but also to obtain information on the tidal friction within the other moons. Most importantly, these future observations may show evidence of Callisto's migration and dissipative processes at play within Jupiter.

Throughout this article, our working hypothesis was the hypothetical fast migration for Callisto proposed by \citet{FULLER-etal_2016}. Yet, even if future measurements prove that the migration of Callisto is slow today, our results show that it may have been fast in the past and still lead to the current configuration of the system. In fact, the dynamical evolutions of all four Galilean moons are deeply coupled, so that the knowledge of all dissipative parameters is essential to reconstruct their orbital history.

\section*{Acknowledgements}
Part of this research has been carried out during a visiting period of G.L. at the IMCCE, whose support is hereby acknowledged. C.G. acknowledges the project MIUR-PRIN 20178CJA2B ``New frontiers of Celestial Mechanics: theory and applications''.

\section*{Data Availability}
The data underlying this article are available in the article.



\bibliographystyle{mnras}
\bibliography{mnrasCallisto} 

\begin{thebibliography}{}
\makeatletter
\relax
\def\mn@urlcharsother{\let\do\@makeother \do\$\do\&\do\#\do\^\do\_\do\%\do\~}
\def\mn@doi{\begingroup\mn@urlcharsother \@ifnextchar [ {\mn@doi@}
  {\mn@doi@[]}}
\def\mn@doi@[#1]#2{\def\@tempa{#1}\ifx\@tempa\@empty \href
  {http://dx.doi.org/#2} {doi:#2}\else \href {http://dx.doi.org/#2} {#1}\fi
  \endgroup}
\def\mn@eprint#1#2{\mn@eprint@#1:#2::\@nil}
\def\mn@eprint@arXiv#1{\href {http://arxiv.org/abs/#1} {{\tt arXiv:#1}}}
\def\mn@eprint@dblp#1{\href {http://dblp.uni-trier.de/rec/bibtex/#1.xml}
  {dblp:#1}}
\def\mn@eprint@#1:#2:#3:#4\@nil{\def\@tempa {#1}\def\@tempb {#2}\def\@tempc
  {#3}\ifx \@tempc \@empty \let \@tempc \@tempb \let \@tempb \@tempa \fi \ifx
  \@tempb \@empty \def\@tempb {arXiv}\fi \@ifundefined
  {mn@eprint@\@tempb}{\@tempb:\@tempc}{\expandafter \expandafter \csname
  mn@eprint@\@tempb\endcsname \expandafter{\@tempc}}}

\bibitem[\protect\citeauthoryear{Batygin \& Morbidelli}{Batygin \&
  Morbidelli}{2020}]{BATYGIN-MORBIDELLI_2020}
Batygin K.,  Morbidelli A.,  2020, \apj, 894, 143

\bibitem[\protect\citeauthoryear{{Celletti}, {Karampotsiou}, {Lhotka},
  {Pucacco}  \& {Volpi}}{{Celletti} et~al.}{2022}]{CELLETTI-etal_2022}
{Celletti} A.,  {Karampotsiou} E.,  {Lhotka} C.,  {Pucacco} G.,   {Volpi} M.,
  2022, Regul. Chaotic Dyn., 27, 381

\bibitem[\protect\citeauthoryear{{Crida}}{{Crida}}{2020}]{CRIDA_2020}
{Crida} A.,  2020, Nat.~Astron., 4, 1024

\bibitem[\protect\citeauthoryear{{Dirkx}, {Gurvits}, {Lainey}, {Lari},
  {Milani}, {Cim{\`o}}, {Bocanegra-Bahamon}  \& {Visser}}{{Dirkx}
  et~al.}{2017}]{DIRKX-etal_2017}
{Dirkx} D.,  {Gurvits} L.~I.,  {Lainey} V.,  {Lari} G.,  {Milani} A.,
  {Cim{\`o}} G.,  {Bocanegra-Bahamon} T.~M.,   {Visser} P.~N.~A.~M.,  2017,
  \planss, 147, 14

\bibitem[\protect\citeauthoryear{{Downey}, {Nimmo}  \& {Matsuyama}}{{Downey}
  et~al.}{2020}]{DOWNEY-etal_2020}
{Downey} B.~G.,  {Nimmo} F.,   {Matsuyama} I.,  2020, \mnras, 499, 40

\bibitem[\protect\citeauthoryear{{Durante} et~al.,}{{Durante}
  et~al.}{2020}]{DURANTE-etal_2020}
{Durante} D.,  et~al., 2020, \grl, 47, e86572

\bibitem[\protect\citeauthoryear{{Efroimsky} \& {Lainey}}{{Efroimsky} \&
  {Lainey}}{2007}]{EFROIMSKY-LAINEY_2007}
{Efroimsky} M.,  {Lainey} V.,  2007, J. Geophys. Res. Planets, 112, E12003

\bibitem[\protect\citeauthoryear{Fuller, Luan  \& Quataert}{Fuller
  et~al.}{2016}]{FULLER-etal_2016}
Fuller J.,  Luan J.,   Quataert E.,  2016, \mnras, 458, 3867

\bibitem[\protect\citeauthoryear{{Goldreich} \& {Soter}}{{Goldreich} \&
  {Soter}}{1966}]{GOLDREICH-SOTER_1966}
{Goldreich} P.,  {Soter} S.,  1966, \icarus, 5, 375

\bibitem[\protect\citeauthoryear{{Greeley}, {Klemaszewski}  \&
  {Wagner}}{{Greeley} et~al.}{2000}]{GREELEY-etal_2000}
{Greeley} R.,  {Klemaszewski} J.~E.,   {Wagner} R.,  2000, \planss, 48, 829

\bibitem[\protect\citeauthoryear{{Greenberg}}{{Greenberg}}{1982}]{GREENBERG_1982}
{Greenberg} R.,  1982, in {Morrison} D.,  ed., , Satellites of Jupiter.
University of Arizona Press, pp 65--92

\bibitem[\protect\citeauthoryear{{Hay}, {Trinh}  \& {Matsuyama}}{{Hay}
  et~al.}{2020}]{HAY-etal_2020}
{Hay} H.~C.~F.~C.,  {Trinh} A.,   {Matsuyama} I.,  2020, \grl, 47, e88317

\bibitem[\protect\citeauthoryear{{Henrard}}{{Henrard}}{1983}]{HENRARD_1983}
{Henrard} J.,  1983, \icarus, 53, 55

\bibitem[\protect\citeauthoryear{{Hussmann} \& {Spohn}}{{Hussmann} \&
  {Spohn}}{2004}]{HUSSMANN-SPOHN_2004}
{Hussmann} H.,  {Spohn} T.,  2004, \icarus, 171, 391

\bibitem[\protect\citeauthoryear{{Kaula}}{{Kaula}}{1964}]{KAULA_1964}
{Kaula} W.~M.,  1964, {Rev. Geophys.}, 2, 661

\bibitem[\protect\citeauthoryear{{Lainey}, {Duriez}  \& {Vienne}}{{Lainey}
  et~al.}{2006}]{LAINEY-etal_2006}
{Lainey} V.,  {Duriez} L.,   {Vienne} A.,  2006, \aap, 456, 783

\bibitem[\protect\citeauthoryear{{Lainey}, {Arlot}, {Karatekin}  \& {van
  Hoolst}}{{Lainey} et~al.}{2009}]{LAINEY-etal_2009}
{Lainey} V.,  {Arlot} J.-E.,  {Karatekin} {\"O}.,   {van Hoolst} T.,  2009,
  \nat, 459, 957

\bibitem[\protect\citeauthoryear{{Lainey} et~al.,}{{Lainey}
  et~al.}{2020}]{LAINEY-etal_2020}
{Lainey} V.,  et~al., 2020, Nat.~Astron., 4, 1053

\bibitem[\protect\citeauthoryear{{Lari}}{{Lari}}{2018}]{LARI_2018}
{Lari} G.,  2018, Celest.~Mech.~Dyn.~Astron., 130, 50

\bibitem[\protect\citeauthoryear{{Lari} \& {Milani}}{{Lari} \&
  {Milani}}{2019}]{LARI-MILANI_2019}
{Lari} G.,  {Milani} A.,  2019, \planss, 176, 104679

\bibitem[\protect\citeauthoryear{{Lari}, {Saillenfest}  \& {Fenucci}}{{Lari}
  et~al.}{2020}]{LARI-etal_2020}
{Lari} G.,  {Saillenfest} M.,   {Fenucci} M.,  2020, \aap, 639, A40

\bibitem[\protect\citeauthoryear{{MacDonald}}{{MacDonald}}{1964}]{MACDONALD_1964}
{MacDonald} G.~J.~F.,  1964, Rev.~Geophys.~Space~Phys., 2, 467

\bibitem[\protect\citeauthoryear{{Madeira}, {Izidoro}  \&
  {Giuliatti~Winter}}{{Madeira} et~al.}{2021}]{MADEIRA-etal_2021}
{Madeira} G.,  {Izidoro} A.,   {Giuliatti~Winter} S.~M.,  2021, \mnras, 504,
  1854

\bibitem[\protect\citeauthoryear{{Malhotra}}{{Malhotra}}{1991}]{MALHOTRA_1991}
{Malhotra} R.,  1991, \icarus, 94, 399

\bibitem[\protect\citeauthoryear{{Mignard}}{{Mignard}}{1979}]{MIGNARD_1979}
{Mignard} F.,  1979, Moon and Planets, 20, 301

\bibitem[\protect\citeauthoryear{{Murray} \& {Dermott}}{{Murray} \&
  {Dermott}}{2000}]{MURRAY-DERMOTT_2000}
{Murray} C.~D.,  {Dermott} S.~F.,  2000, {Solar System Dynamics}.
Cambridge University Press

\bibitem[\protect\citeauthoryear{Paita, Celletti  \& Pucacco}{Paita
  et~al.}{2018}]{PAITA-etal_2018}
Paita F.,  Celletti A.,   Pucacco G.,  2018, \aap, 617, A35

\bibitem[\protect\citeauthoryear{{Peale} \& {Lee}}{{Peale} \&
  {Lee}}{2002}]{PEALE-LEE_2002}
{Peale} S.~J.,  {Lee} M.~H.,  2002, Science, 298, 593

\bibitem[\protect\citeauthoryear{{Schubert}, {Anderson}, {Spohn}  \&
  {McKinnon}}{{Schubert} et~al.}{2004}]{SCHUBERT-etal_2004}
{Schubert} G.,  {Anderson} J.~D.,  {Spohn} T.,   {McKinnon} W.~B.,  2004, in
  {Bagenal} F.,  {Dowling} T.~E.,   {McKinnon} W.~B.,  eds, , Jupiter. The
  planet, satellites and magnetosphere.
Cambridge University Press, Chapt.~13, pp 281--306

\bibitem[\protect\citeauthoryear{{Shibaike}, {Ormel}, {Ida}, {Okuzumi}  \&
  {Sasaki}}{{Shibaike} et~al.}{2019}]{SHIBAIKE-etal_2019}
{Shibaike} Y.,  {Ormel} C.~W.,  {Ida} S.,  {Okuzumi} S.,   {Sasaki} T.,  2019,
  \apj, 885, 79

\bibitem[\protect\citeauthoryear{Showman \& Malhotra}{Showman \&
  Malhotra}{1997}]{SHOWMAN-MALHOTRA_1997}
Showman A.~P.,  Malhotra R.,  1997, \icarus, 127, 93

\bibitem[\protect\citeauthoryear{{Sinclair}}{{Sinclair}}{1975}]{SINCLAIR_1975}
{Sinclair} A.~T.,  1975, Celest. Mech., 12, 89

\bibitem[\protect\citeauthoryear{{Tittemore}}{{Tittemore}}{1990}]{TITTEMORE_1990}
{Tittemore} W.~C.,  1990, Science, 250, 263

\bibitem[\protect\citeauthoryear{{Tittemore} \& {Wisdom}}{{Tittemore} \&
  {Wisdom}}{1988}]{TITTEMORE-WISDOM_1988}
{Tittemore} W.~C.,  {Wisdom} J.,  1988, \icarus, 74, 172

\bibitem[\protect\citeauthoryear{{Veeder}, {Matson}, {Johnson}, {Blaney}  \&
  {Goguen}}{{Veeder} et~al.}{1994}]{VEEDER-etal_1994}
{Veeder} G.~J.,  {Matson} D.~L.,  {Johnson} T.~V.,  {Blaney} D.~L.,   {Goguen}
  J.~D.,  1994, J. Geophys. Res. Planets, 99, 17095

\bibitem[\protect\citeauthoryear{{Wahl}, {Parisi}, {Folkner}, {Hubbard}  \&
  {Militzer}}{{Wahl} et~al.}{2020}]{WAHL-etal_2020}
{Wahl} S.~M.,  {Parisi} M.,  {Folkner} W.~M.,  {Hubbard} W.~B.,   {Militzer}
  B.,  2020, \apj, 891, 42

\bibitem[\protect\citeauthoryear{Yoder}{Yoder}{1979}]{YODER_1979}
Yoder C.~F.,  1979, \nat, 279, 767

\bibitem[\protect\citeauthoryear{{Yoder} \& {Peale}}{{Yoder} \&
  {Peale}}{1981}]{YODER-PEALE_1981}
{Yoder} C.~F.,  {Peale} S.~J.,  1981, \icarus, 47, 1

\makeatother
\end{thebibliography}


\bsp	
\label{lastpage}
\end{document}